\documentclass{amsart}
\usepackage{amssymb}
\usepackage{amsfonts}
\usepackage{graphicx}
\usepackage{epstopdf}
\usepackage{hyperref}
\numberwithin{equation}{section}

\newcommand{\beq}{\begin{equation}}
\newcommand{\eeq}{\end{equation}}

\begin{document}

\begin{center}
{\Large \bf Nonperturbative universal Chern-Simons theory} \\
\vspace*{1 cm}
{\large  R.L.Mkrtchyan 
}
\vspace*{0.5 cm}

{\small\it Yerevan Physics Institute, 2 Alikhanian Br. Str., 0036 Yerevan, Armenia}

{\small\it E-mail: mrl55@list.ru}

\end{center}\vspace{2cm}

{\small  {\bf Abstract.} Closed simple integral representation through Vogel's universal parameters is found both for perturbative and nonperturbative (which is inverse invariant group volume) parts of free energy of Chern-Simons theory on $S^3$. This proves  the universality of that partition function. For classical groups it manifestly satisfy $N \rightarrow - N$ duality, in apparent contradiction with previously used ones. For $SU(N)$ we show that asymptotic of nonperturbative part of our partition function  coincides with that of Barnes G-function, recover Chern-Simons/topological string duality in genus expansion and resolve abovementioned contradiction. We discuss  few possible directions of development of these results: derivation of representation of free energy through Gopakumar-Vafa invariants, possible appearance of non-perturbative additional terms,  $1/N$ expansion for exceptional groups, duality between string coupling constant and K\"ahler parameters, etc.}

\section{Introduction}

Many theories require for their definition  Lie algebra, particularly simple Lie algebra. Due to well-known classification of these algebras this choice  is considered as discrete one, although  properties of theories with different simple Lie algebras often are expected to be similar. It was 't Hooft  \cite{H1} who first promotes the integer parameter $N$ of $SU(N)$ gauge group into continuous one and suggests an extremely  fruitful idea of $1/N$ expansion of gauge theories, connecting them with strings theory. This idea naturally extends to other classical groups - $SO(N)$ and $Sp(2N)$. Of course, it requires some continuation of theories from the domain of integer N to entire real line and complex plane. Even assuming analyticity of functions this extension is not unique. Possible (actually used in \cite{H1}) definition is based on a fact that gauge theory perturbation terms can be presented as  polynomial (or rational) functions of $N$, and in this form they are extended to an arbitrary $N$. Further, it appears that $SO(2N)$ and $Sp(2N)$ gauge theories are connected by $N\rightarrow -N$ transformation  \cite{Mkr,Cvitbook} ($SU(N)$ is self-dual w.r.t. the $N\rightarrow -N$). So, in this way gauge theories with classical groups are promoted into domain of arbitrary N, and joined into two objects instead of four classical series. But gauge theories with exceptional groups remain completely out of this extension and unification, and the question whether they can be obtained  in a similar way as a discrete points on some line (or other manifold), from which physical quantities depend smoothly, and whether that manifold can include classical groups, remains completely open until recently. 

An important development happens in  \cite{MV1}, on the basis of Vogel's approach to knot theory Vassiliev's invariants \cite{V0,V}. In these works Vogel sought the most general weight system for these invariants. Any simple Lie algebra provides such a weight system, question was whether one can find more general one. It was proved that one-index contraction of two vertexes (vertex is a structure constants in case of Lie algebra), which can be considered as an operator acting on a symmetric square of corresponding space (adjoint representation) satisfies a third order equation (after factorization over trivial one-dimensional eigenstate). Then one can try to parametrize weight systems by three parameters, eigenvalues of that operator. According to appearance of these parameters, theory is symmetric w.r.t.   permutations of them. Also an overall factor is inessential, since it corresponds to change in  normalization of invariant bilinear form on algebra. So these three Vogel parameters $\alpha,\beta, \gamma$ belong to the Vogel's plane which is by definition a projective plane factorized w.r.t. all permutations of projective parameters. It was calculated in \cite{V0} what are specific points in that plane simple Lie algebras correspond to. They are  given in Table \ref{tab:1}. Parameter $t$ there denotes the sum $t=\alpha+\beta+\gamma$, last column shows that with normalization of first column $t$ becomes a dual Coxeter number $h^\vee$. This normalization corresponds to the so called minimal bilinear invariant form on the simple Lie algebra, defined to have the square of long root(s) equal 2.

\begin{table}[h] \label{tab:1}
\caption{Vogel's parameters for simple Lie algebras}     
\begin{tabular}{|c|c|c|c|c|c|}
\hline
Type & Lie algebra  & $\alpha$ & $\beta$ & $\gamma$  & $t=h^\vee$\\   
\hline    
$A_n$ &  $\mathfrak {sl}_{n+1}$     & $-2$ & 2 & $(n+1) $ & $n+1$\\
$B_n$ &   $\mathfrak {so}_{2n+1}$    & $-2$ & 4& $2n-3 $ & $2n-1$\\
$C_n$ & $ \mathfrak {sp}_{2n}$    & $-2$ & 1 & $n+2 $ & $n+1$\\
$D_n$ &   $\mathfrak {so}_{2n}$    & $-2$ & 4 & $2n-4$ & $2n-2$\\
$G_2$ &  $\mathfrak {g}_{2}  $    & $-2$ & $10/3 $& $8/3$ & $4$ \\
$F_4$ & $\mathfrak {f}_{4}  $    & $-2$ & $ 5$& $ 6$ & $9$\\
$E_6$ &  $\mathfrak {e}_{6}  $    & $-2$ & $ 6$& $ 8$ & $12$\\
$E_7$ & $\mathfrak {e}_{7}  $    & $-2$ & $ 8$& $ 12$ & $18$ \\
$E_8$ & $\mathfrak {e}_{8}  $    & $-2$ & $ 12$& $20$ & $30$\\
\hline  
\end{tabular}
\end{table}

Next, many quantities can be expressed through Vogel universal parameters by smooth functions. Such  quantities are called universal. For example, dimension of adjoint representation and its second Casimir are universal quantities, since they are given by 

\begin{eqnarray}
\label{f3}
dim \, \mathfrak {g} &=& \frac{(\alpha-2t)(\beta-2t)(\gamma-2t)}{\alpha\beta\gamma}\\
C_2&=&2t
\end{eqnarray}
Dimensions naturally are integers for parameters from Table \ref{tab:1}, although not only for them. In \cite{Del,DM,LM1,LM2,LM4} a lot of dimension formulae have been presented, covering many irreducible representations of simple Lie algebras. In \cite{MSV} the universal formula is presented for eigenvalues of higher Casimirs on an adjoint representation.  In ref \cite{Mkr2} the classification of simple Lie algebras is recovered, in a reasonable sense, by requirement of regularity of universal character (\ref{gene} below) and it is shown that simple Lie algebras (and few similar objects) are in one to one correspondence with solutions of certain Diophantine equations. 

But it is not guaranteed, that any quantity can be expressed through universal parameters. For example the universal formula for dimension of fundamental representations of classical groups is not known (although they are connected by $N\rightarrow -N$ duality). They are "not universal" in this sense, although it is not a theorem, and there is no clear definition of irrep being "not universal". It is not excluded that corresponding formulae exist, but are more complicated. 

The $N\rightarrow -N$ duality is overlapping with universality. If some quantity is universal, then for classical groups it is  $N\rightarrow -N$ dual: as is seen from Table \ref{tab:1} that duality corresponds to a $\alpha \leftrightarrow \beta$ transposition together with projective transformation. For that quantities universality is wider notion, since it includes an exceptional groups. From the other hand, some quantities, which are duality related (as mentioned dimensions of fundamental representations of classical groups) have no known universal representation. 
 
Idea of \cite{MV1} was, independently of the final conclusions of works \cite{V0,V,Del}(which actually are on the stage of some reasonable conjectures), to try to use this approach to extend ideas of $1/N$ expansion to gauge theories with an arbitrary simple Lie algebras. The main question  is whether physical theories, particularly gauge theories, "respect" Vogel's parameters, i.e. whether physical quantities can be expressed through these parameters by some smooth functions, which give answers for particular group at values from Table \ref{tab:1}.   The main achievement of \cite{MV1} is that it is proved for Chern-Simons gauge theory on $S^3$  that few important quantities are universal. Particularly, perturbative part of its partition function is universal, i.e. can be expressed through universal parameters, more exactly, each term in perturbative expansion is a rational function of universal parameters.  Universal are also central charge and unknot Wilson loop in adjoint representation. 

In present paper we have developed \cite{MV1} in two directions. First, we present a closed universal integral representation for perturbative part $F_2$ of free energy of Chern-Simons on $S^3$: 
\begin{eqnarray} \label{1}
F_2=\int^{\infty}_0 dx \frac{F(x/\delta)}{x(e^{x}-1)} 
\end{eqnarray}
Here $F(x)$ is a universal expression \cite{MV1} for character of adjoint representation minus dimension of algebra, evaluated at point $x \rho$, $\rho$ is a Weyl vector in roots space, i.e. half of sum of all positive roots of a given simple Lie algebra.

\begin{eqnarray}\label{gene}
F(x)&=&\frac{\sinh(x\frac{\alpha-2t}{4})}{\sinh(\frac{x\alpha}{4})}\frac{\sinh(x\frac{\beta-2t}{4})}{\sinh(x\frac{\beta}{4})}\frac{\sinh(x\frac{\gamma-2t}{4})}{\sinh(x\frac{\gamma}{4})}-\frac{(\alpha-2t)(\beta-2t)(\gamma-2t)}{\alpha\beta\gamma}\\
\delta&=&\kappa+t
\end{eqnarray}
$\kappa$ is a coefficient in front of Chern-Simons action (more detailed definitions see in Section 2), $\delta$ is (an inverse) effective coupling constant.

Second, it is argued that the same expression (with minus sign)  with  $\kappa=0$  essentially gives a non-perturbative part of free energy (exact expression is given by (\ref{Z1}),(\ref{vol}),(\ref{totalfree})). This particularly means that non-perturbative part is  universal, also,  which completes the proof of universality of total partition function of Chern-Simons theory on $S^3$. Since non-perturbative part of partition function on $S^3$ is essentially an (inverse) invariant volume of gauge group, it means that we have found a universal expression for invariant volume of group, see (\ref{vol}). Its logarithm is essentially (\ref{1}) with $\kappa=0$. Note the beautiful expression for $1/N$ expansion of $SU(N)$ group's volume given in \cite{OV}, where coefficients are virtual Euler characteristics of moduli space of curves of genus $g$. The generalization of this expression is one of the aims of our search of universal expression for volume. Particularly, when expanded over universal parameters, universal volume will provide some refinement of abovementioned virtual Euler characteristics, the topological interpretation of which is very interesting question, with possible  influence on string theory. 

This representation of free energy of Chern-Simons theory apparently differs  from known answers  for classical groups, particularly $SU(N)$. For perturbative part it is more or less evident, by construction, that when expanded into a series over coupling constant (actually over $1/\delta\sim 1/(k+N)$) it will give the same answers, polynomials over $N$. But for non-perturbative part one cannot be sure at all, because we are dealing with complicated function of $N$, which essentially is product over $n$ of $n!$, $n$ running from 1 to $N-1$. Usually this is continued to arbitrary complex $N$ as a Barnes function $G(1+N)$, see \cite{Pe,GV,OV}, but since it is not unique one may wonder what is the relation between these two functions: Barnes (left side of Eq.\ref{GN}) and our one (right side of Eq.\ref{GN}). Moreover, usual continuation violates $N\rightarrow -N$ duality (Barnes function $G(1+N)$ is not even with respect to $N$, recovering this duality was another our aim), which is one of the questions  we had to resolve when asking for a universal expression for free energy, because, as mentioned above, $N\rightarrow -N$ duality is part of universality. Indeed, expressions we obtained for $SU(N)$ partition function's perturbative and non-perturbative parts  (\ref{genesun}), (\ref{volsun}), (\ref{GN}) are explicitly $N\rightarrow -N$ invariant, so we have an explicitly different, from the usual one, expression for non-perturbative part of $SU(N)$ Chern-Simons. 

To resolve this puzzle, establish a connection with previous formulae for partition function,  and also to work out technique of handling the integral representations of free energy, we calculate a genus expansion of free energy of $SU(N)$ theory and compare with that of \cite{GV} (see \cite{Mar1} for a review). I.e. we re-derive in our language of integral representations the proof of duality of Chern-Simons on $S^3$ with closed topological strings, known as a Gopakumar-Vafa  geometrical transition. We show that these genus expansions coincide, particularly   genus expansion of (essentially, few details have to be taken into account)  Barnes function coincides with genus expansion of our non-perturbative universal expression for an $SU(N)$ group volume.  This establishes the coincidence of both representations of free energy of Chern-Simons theory for $SU(N)$. Advantage of  universal one is $N\rightarrow -N$ duality for $SU(N)$ (and $SO(2N)/Sp(2N)$). This also is a strong argument in favor of universal expression for free energy.  We also explain apparent contradiction described above in Section \ref{sectbarn}.

In Conclusion we discuss few topics for future development. Among them are expression of free energy through Gopakumar-Vafa invariants - how it can appear in the present approach, and what are additional non-perturbative terms; $1/N$ expansion for exceptional groups and refinement of topological invariants; possible duality between string coupling constant and K\"ahler parameters, universality of partition function for other manifolds, etc.

\section{Perturbative partition function of Chern-Simons theory}

Partition function of Chern-Simons theory is formally given by functional integral
\begin{eqnarray}
Z(M)=\int DA \, exp\left(\frac{i\kappa}{4\pi}\int_M Tr \left( A \wedge d A + \frac{2}{3} A\wedge A \wedge A \right)\right)
\end{eqnarray}
Here "Tr" means an unnormalized invariant bilinear form on algebra, $\kappa $ is an arbitrary number, which has to become a positive integer $k$ when "Tr" is normalized to  minimal invariant bilinear form on algebra, see \cite{MV1}. So $\kappa$ has to be rescaled simultaneously with universal parameters. The values of universal parameters, corresponding to minimal invariant bilinear form are given in Table \ref{tab:1}, then $\kappa$ becomes an integer, level $k$, parameter $t$ becomes dual Coxeter number $h^{\vee}$. 

Partition function of Chern-Simons theory on $S^3$, with a specific choice of trivialization of tangent bundle, is \cite{W1};

\begin{eqnarray}
\label{zz*}
Z(S^3) = S_{00}=Vol(Q^\vee)^{-1}\delta^{-r/2}\prod_{\mu \in R_+}
2\sin \frac{\pi (\mu,\rho)}{\delta}.
\end{eqnarray}
where $S_{00}$ is an element of matrix of modular S-transformations,  $R (R_+)$ is the set of roots (positive roots) of algebra, $Q^\vee$ is coroot lattice,  $Vol(Q^\vee)$ is a volume of fundamental domain of that lattice, $\delta=\kappa+t$, product is over all positive roots of a given algebra. 

From now on we shall omit argument $S^3$ of partition function, since below we consider only $S^3$ manifold.  Let's rewrite it as the product $Z=Z_{1} Z_2,$ where
\begin{eqnarray}
\label{zz1}
Z_1=Vol(Q^\vee)^{-1}\delta^{-r/2}\prod_{\mu \in R_+}
\frac{2\pi (\mu,\rho)}{\delta}
\end{eqnarray}

and

\begin{eqnarray}
\label{zz2}
Z_2=\prod_{\mu \in R_+}
\sin \frac{\pi (\mu,\rho)}{\delta}/\frac{\pi (\mu,\rho)}{\delta}.
\end{eqnarray}

The first, non-perturbative factor $Z_1$ has a geometric meaning (cf. \cite{W1,OV,MV1}):

\begin{eqnarray}
\label{zz12}
Z_1=\frac{(2\pi  \delta^{-1/2})^{dim \, \mathfrak g}} {Vol(G)},
\end{eqnarray}
where $Vol(G)$ is the volume of the corresponding compact simple simply connected group, with metric induced from the minimal bilinear invariant form on the algebra.

This factor will be discussed in next Section, here we shall calculate $Z_2$, the  perturbative part of free energy.
Using infinite product representation
\begin{eqnarray}
\sin \pi x= \pi x \prod^{\infty}_{n=1} (1-\left(\frac{x}{n}\right)^2)
\end{eqnarray}
and 
\begin{eqnarray}
\ln \frac {\sin \pi x}{\pi x}= \sum ^{\infty}_{n=1} \ln \left(1-\left(\frac{x}{n}\right)^2\right)=-\sum ^{\infty}_{n=1} \sum ^{\infty}_{m=1} \frac{1}{m}\frac{x^{2m}}{n^{2m}}
\end{eqnarray}
we get for perturbative free energy:

\begin{eqnarray}\label{fr}
F_2=-\ln Z_2=\sum_{\mu \in R_+}\sum ^{\infty}_{n=1} \sum ^{\infty}_{m=1}\frac{1}{m}  \frac{\left(  \frac{(\mu,\rho)}{\delta} \right) ^{2m}}{n^{2m}}=\\
=\sum ^{\infty}_{n=1} \sum ^{\infty}_{m=1}\frac{1}{2m}  \frac{ p_{2m} }{(n\delta)^{2m}}
\end{eqnarray}

where we introduced

$$p_k=\sum_{\mu \in R} (\mu,\rho)^{k}.$$
We have $p_k=0$ for all odd $k$ and $p_{2m}=2 \sum_{\mu \in R_+} (\mu,\rho)^{2m}$. Up to now, calculations are standard. Next, consider  following exponential generating function, which evidently is character in adjoint representation, evaluated at point $x\rho$, minus dimension of algebra \cite{MV1}:
$$F(x)=\sum_{k=1}^{\infty}\frac{p_{k}}{k!} x^{k}=
\sum_{\mu\in R} (e^{x(\mu,\rho)}-1).$$
It was expressed in terms of the Vogel's parameters in \cite{MV1}, answer is (\ref{gene}), which shows that $p_k$ and hence perturbative partition function are universal quantities. Expression for $p_2$ is a homogeneous form of the so called  Freudenthal-de Vries strange formula:

\begin{eqnarray}
\sum_{\mu \in R_+} (\mu, \rho)^{2}= \frac{t^2}{12} dim \, \mathfrak {g}, 
\end{eqnarray}

Expressions for other $p_k$ can be called a generalized Freudenthal-de Vries (strange) relations.

Now we  transform (\ref{fr}) further by  Borel summation of m-series in it. Answer can be expressed through $F(x)$:
\begin{eqnarray}
\label{f2}
F_2=\sum^{\infty}_{n=1} \int^{\infty}_0 dx \frac{e^{-x}}{x} F(\frac{x}{n\delta})=\sum^{\infty}_{n=1} \int^{\infty}_0 dx \frac{e^{-xn\delta}}{x} F(x)=\\
\int^{\infty}_0 dx \frac{e^{-x\delta}}{x(1-e^{-x\delta})} F(x)=
\int^{\infty}_0 \frac{dx}{x} \frac{F(x/\delta)}{(e^{x}-1)} 
\end{eqnarray}

Integrals converge, since at large $x$ in physical region of parameters $F(x/\delta)$ behaves as exponent of $x$ with index $\frac{t+\alpha/2}{\delta}$, which is always less than 1. Physical region of parameters we define as (up to sign) $ \alpha<0, \beta>0, \gamma>0, t>0, \kappa >0$,   to which all really existing Chern-Simons theories with simple Lie algebras belong.  At $x=0$ integral converges due to $F(x)\sim x^2$ at small $x$. 

Of course, recovering a function from its expansion (most probably asymptotic one) is not unique, so we provide below another, straightforward and simple  derivation of (\ref{f2}). That derivation requires certain integral representation of gamma-function. 

Recall representation of $sin$ through gamma-functions:
\begin{eqnarray}
\frac{sin(\pi x)}{\pi x}=\frac{1}{\Gamma(1-x) \Gamma(1+x)}
\end{eqnarray}

Then perturbative free energy is:

\begin{eqnarray}
F_2=-\sum_{\mu\in R_+}\ln(\frac{sin(\pi (\rho,\mu)/\delta)}{\pi (\rho,\mu)/\delta)})\\
=\sum_{\mu\in R_+}\ln(\Gamma(1-(\rho,\mu)/\delta))+\ln(\Gamma(1+(\rho,\mu)/\delta))
\end{eqnarray}

Next we use integral representation of gamma-function \cite{gamma}:
\begin{eqnarray}
\ln\Gamma(1+z)=\int_{0}^{\infty}dx e^{-x}\frac{e^{-zx}+z(1-e^{-x})-1}{x(1-e^{-x})}
\end{eqnarray}
to get
\begin{eqnarray}
F_2=
\int_{0}^{\infty}dx\frac{e^{-x}\sum_{\mu\in R_+} \left(e^{x\frac{(\rho,\mu)}{\delta}}+e^{-x\frac{(\rho,\mu)}{\delta}}-2\right)}{x(1-e^{-x})}=\\
\int^{\infty}_0 dx \frac{e^{-x}}{x(1-e^{-x})} F(x/\delta)
\end{eqnarray}
which coincides with above.

Universal perturbative partition function expression specified for $SU(N)$ gives 
(taking parameters in normalization from Table \ref{tab:1}: $(\alpha, \beta, \gamma, \kappa)=(-2,2,N,k)$):

\begin{eqnarray}\label{genesun}
\int^{\infty}_0 dx \frac{e^{-x}}{x(1-e^{-x})} (\frac{\sinh(x\frac{N+1}{2(k+N)})\sinh(x\frac{N-1}{2(k+N)})}{\sinh^2(\frac{x}{2(k+N)})}-(N^2-1))=\\
\int^{\infty}_0 dx \frac{e^{-x}}{x(1-e^{-x})} \left(\frac{\cosh(\frac{Nx}{k+N})-\cosh(\frac{x}{k+N})}{2\sinh^2(\frac{x}{2(k+N)})}-(N^2-1)\right)=\\
\int^{\infty}_0 dx \frac{e^{-x}}{x(1-e^{-x})} \left(\frac{\cosh(\frac{Nx}{k+N})-1}{2\sinh^2(\frac{x}{2(k+N)})}-N^2\right)
\end{eqnarray}
which is explicitly invariant w.r.t. the duality $N \rightarrow -N, k \rightarrow -k$. 

For $SO(N)$ we have:
\begin{eqnarray}
\int^{\infty}_0 dx \frac{e^{-x}}{x(1-e^{-x})} F(\frac{x}{\delta})\\
 F(\frac{x}{\delta})=\frac{\cosh(\frac{x(N-4)}{4(k+N-2)})\sinh(\frac{x(N-1)}{2(k+N-2)})\sinh(\frac{Nx}{4(k+N-2)})}{\sinh(\frac{x}{2(k+N-2)})\sinh(\frac{x}{k+N-2})}-\frac{N(N-1)}{2}
\end{eqnarray}

Calculation for $Sp(2N)$ gives the same answer as for $SO(2N)$ with $N \rightarrow - N, k \rightarrow -2k$, as it should be according to $N \rightarrow - N$ duality of Chern-Simons \cite{MV1}.

\section{Non-perturbative part of partition function}

Partition function of Chern-Simons theory on $S^3$, as defined in \cite{W1}, is equal to 1 at $\kappa=0$. This  can be checked explicitly on exact expressions, e.g. for $SU(N)$, in general it follows  from the fact that an $S_{00}$ element of modular transformations matrix  at $\kappa=0$ is unity since there is no nontrivial unitary representation of corresponding affine algebra. So, nonperturbative part of partition function can be expressed through perturbative (see (\ref{Z1}) below) and is universal since latter one is universal. The same is right about volume of group, see (\ref{vol}).

Partition function is product of non-perturbative $Z_1$ and perturbative $Z_2$ parts:

\begin{eqnarray}
Z=Z_1Z_2 
\end{eqnarray}

Perturbative part depends on $\kappa$: $Z_2=Z_2(\kappa)$, dependence of $Z_1$ on $\kappa$ is given by power of $\delta$ (\ref{zz12}):

\begin{eqnarray}
Z_1(\kappa)=Vol(G)^{-1} ((2\pi\delta^{-1/2})^{dim \, \mathfrak {g}}
\end{eqnarray}

At  $\kappa =0$ one have $\delta=t$ and $Z=1$, as discussed above, so 

\begin{eqnarray} \label{Z1}
Z_1(\kappa)=\frac{1}{Z_2(0)} \left( \frac{t}{\delta}\right)^{(dim \, \mathfrak {g})/2} \\ \label{vol}
Vol(G)=Z_2(0)(2\pi t^{-1/2})^{dim \, \mathfrak {g}}
\end{eqnarray}

This gives a universal integral representation for volume function. Direct derivation of this expression, without use of partition function of Chern-Simons theory, can be given with the use of formulae for volume given in \cite{KP}, applying the same transformations.

So, partition function can be expressed as: 

\begin{eqnarray}
Z=\frac{Z_2(\kappa)}{Z_2(0)} \left( \frac{t}{\delta}\right)^{dim/2} 
\end{eqnarray}

and final form of complete free energy is

 \begin{eqnarray}\label{totalfree}
 F_1+F_2=(dim/2)\ln(\delta/t) + \int^{\infty}_0 \frac{dx}{x} \frac{F(x/\delta)-F(x/t)}{(e^{x}-1)} 
 \end{eqnarray}

Logarithm of main multiplier of volume of $SU(N)$, given by $Z_2(\kappa)$ at  $\kappa=0$, is:

\begin{eqnarray} \label{volsun}
\int^{\infty}_0 dx \frac{1}{x(e^{x}-1)} \left(\frac{\cosh(x)-\cosh(x/(N))}{2\sinh^2(\frac{x}{2N})}-(N^2-1)\right)=\\
\int^{\infty}_0 dx \frac{1}{x(e^{x}-1)} \left(\frac{\cosh(x)-1}{2\sinh^2(\frac{x}{2N})}-N^2\right)=\\
\int^{\infty}_0 dx \frac{1}{x(e^{x}-1)} \left(\frac{\sinh^2(\frac{x}{2})}{\sinh^2(\frac{x}{2N})}-N^2\right)=\\
\int^{\infty}_0 \frac{dx}{x}\left(\frac{1-e^{-x}}{4\sinh^2(\frac{x}{2N})}-\frac{N^2}{e^x-1}\right)
\end{eqnarray}

\section{Topological strings and $1/N$ expansion of $SU(N)$ Chern-Simons} \label{secsun}

Expressions for free energy, obtained in previous sections, in  cases of classical groups should recover the known answers for a  $1/\delta, 1/N$ series, which are shown to coincide with those for closed topological strings on a manifold with changed topology (Gopakumar-Vafa geometrical transition \cite{GV,GV1,GV2,OV}). We shall consider $SU(N)$ case and obtain all terms in these expansions, which appear to coincide with known ones. We particularly present an integral representations for them, using a known  integral representation for polylogarithm and other functions. This will be a non-trivial check for our formulae, particularly for universal expression for non-perturbative free energy of Chern-Simons theory. Let's stress again, that analytic continuation from integer $N$ (at which gauge theory was defined initially) to an arbitrary complex values is not unique, so coincidence of our expressions and those using Barnes function is not guaranteed and should be checked, in this or that way. Here we establish a coincidence of asymptotic expansions, important for duality with topological strings. This, however, is not sufficient for statement of coincidence of Barnes function and volume function (with corresponding details, see Section \ref{sectbarn}), which is actually subsequently proved in \cite{Vv} by direct transformation of integral representation.  

With the use of identities
\begin{eqnarray}
\partial \coth\frac{x}{2}=-\frac{1}{2\sinh^2\frac{x}{2}}\\
\cosh(x)=2\sinh^2(\frac{x}{2})+1\\
\coth\frac{x}{2}=\sum_{n=0}^{\infty}\frac{2B_{2n}x^{2n-1}}{(2n)!}=
\frac{2}{x}+\frac{x}{6}-\frac{x^3}{8\cdot 45}+...
\end{eqnarray}

one can get an expansion of $F(x)$ for $SU(N)$ over x:

\begin{eqnarray}\label{expF}
F(x)=(\frac{\cosh(Nx))-\cosh(x)}{2\sinh^2(\frac{x}{2})}-(N^2-1))=\\
-1+\partial \coth\frac{x}{2}-\cosh(Nx)\partial \coth\frac{x}{2}-(N^2-1)=\\
\label{fsun}
-\sum_{g=0,l=1}^{\infty}\frac{2(2g-1)}{(2g)!(2l)!}B_{2g}N^{2l}x^{2l+2g-2}-N^2
\end{eqnarray}

To get elements of perturbative expansion of free energy, we have to multiply each term on $(2l+2g-2)!/2(l+g-1)$, substitute $x$ by $1/(\delta n)$ and sum over $n$. All this happens inside our integral representation of free energy, if we insert in it the expansion of F(x). To get a complete answer, with the purpose of establish a duality with topological strings, one have to add a terms from non-perturbative part, expanded in $1/N$. Below we shall recover the known proof of duality and show how it reappear in our language of integral representations of different contributions into free energy. These are the calculations of Gopakumar-Vafa \cite{GV}, we shall compare with presentation of Mari\~no \cite{Mar1}. 

So, to make a contact with known results we multiply each term of (\ref{fsun}) on $(2l+2g-2)!/2(l+g-1)$, substitute $x$ by $1/(\delta n)$ and sum over $n$. Also, introduce string coupling constant $g_s$ and 't Hooft coupling (K\"ahler parameter) $\mu$:
\begin{eqnarray}
g_s=\frac{2\pi}{k+N}, \mu=\frac{2\pi iN}{k+N}=ig_sN
\end{eqnarray}  

Then for $ g\geq 2 $:

\begin{eqnarray}
\sum_{n=1}^{\infty}\frac{2(2g-1)}{(2g)!(2l)!}B_{2g}N^{2l} \left(\frac{1}{\delta n}\right)^{2l+2g-2}\frac{(2l+2g-2)!}{2(l+g-1)}=\\
 -\frac{2B_{2g}\zeta(2l+2g-2)}{2g(2g-2)}C_{2l}^{2l+2g-3}\frac{N^{2l}}{\delta^{2l+2g-2}}=\\
-\frac{2B_{2g}\zeta(2l+2g-2)}{2g(2g-2)(2\pi)^{2g+2l-2}}C_{2l}^{2l+2g-3}N^{2l}(g_s)^{2l+2g-2}
\end{eqnarray}

which coincides essentially with (3.16), (3.18) of  \cite{Mar1}. The same is true for terms with $g=0,1$, they coincide with (3.17) of \cite{Mar1}.

Going back to our integral representation, insert into that of $F_2$ the expansion of character \ref{expF}, and consider particular terms with fixed $g \geq 2$:

\begin{eqnarray}
\int^{\infty}_0 dx \frac{e^{-x}}{x(1-e^{-x})}(-1)\sum_{l=1}^{\infty}\frac{2(2g-1)}{(2g)!(2l)!}B_{2g}N^{2l}\frac{x^{2l+2g-2}}{\delta^{2l+2g-2}}=\\
\int^{\infty}_0 dx \frac{e^{-x}}{x(1-e^{-x})}(-1)\frac{2(2g-1)B_{2g}}{(2g)!}\left(\frac{x}{\delta}\right)^{2g-2}\left(\cosh\frac{Nx}{\delta}-1\right)
\end{eqnarray}

Take one of two terms of last expression, namely last term, -1, in the last bracket:
\begin{eqnarray}
\int^{\infty}_0 dx \frac{e^{-x}}{x(1-e^{-x})}(-1)\frac{2(2g-1)B_{2g}}{(2g)!}\left(\frac{x}{\delta}\right)^{2g-2}(-1)=\\
\left(\frac{1}{\delta}\right)^{2g-2}\frac{2(2g-1)B_{2g}}{(2g)!} \int^{\infty}_0 dx x^{2g-3}(e^{-x}+e^{-2x}+e^{-3}+...)=\\
\left(\frac{1}{\delta}\right)^{2g-2}\frac{2(2g-1)B_{2g}}{(2g)!}(2g-3)!(1+\frac{1}{2^{2g-2}}+\frac{1}{3^{2g-2}}+...)=\\
\left(\frac{1}{\delta}\right)^{2g-2}\frac{2(2g-1)B_{2g}}{(2g)!}(2g-3)!\zeta(2g-2)=\\
\left(\frac{1}{\delta}\right)^{2g-2}\frac{2B_{2g}}{(2g-2)(2g)}  (-1)^{g}\frac{B_{2g-2}(2\pi)^{2g-2}}{2(2g-2)!}=\\
g_s^{2g-2}\frac{(-1)^{g} B_{2g}B_{2g-2}}{(2g-2)(2g)(2g-2)!} \label{BB}
\end{eqnarray}

where we use

\begin{eqnarray}
\zeta(2m)=(-1)^{m+1}\frac{B_{2m}(2\pi)^{2m}}{2(2m)!}
\end{eqnarray}

This is exactly the  first terms in (5.28) and (5.32) of \cite{Mar1}, up to sign, which comes from different definitions of free energy. It remains to obtain the last term in (5.32), i.e. that with polylogarithms. In \cite{Mar1} it is obtained as a sum of remaining perturbative terms in (5.28), which corresponds to term with $cosh$ above, and terms from non-perturbative $Z_1$ part above. So, we have to sum up that $cosh$ term and term of order $N^{2-2g}$ from asymptotic expansion of group volume (more exactly, $Z_1$) term. Let's find that asymptotic expansion, later on we shall compare that with the expansion of Barnes $G$-function \cite{Barnes1,Mar1,OV}.

Nonperturbative contribution to free energy is (below we explicitly write dependence of free energies $F_1(k), F_2(k)$ from parameter $k$)
\begin{eqnarray}
F_1(k)=-\ln Z_1=\frac{1}{2}(N^2-1)\ln \frac{k+N}{N}-F_2(0)
\end{eqnarray}

 where $F_2(0)$ is:
\begin{eqnarray}
F_2(0)=\int^{\infty}_0 \frac{dx}{x}\left(\frac{1-e^{-x}}{4\sinh^2(\frac{x}{2N})}-\frac{N^2}{e^x-1}\right)
\end{eqnarray}

Integral is convergent, but to consider different contributions separately, let's regularize it at low limit by small $\epsilon$, calculate contributions up to $O(\epsilon)$, then sum all contributions and send $\epsilon \rightarrow 0$. First contribution is: 

\begin{eqnarray}
\int^{\infty}_{\epsilon} \frac{dx}{x}\frac{1}{4\sinh^2(\frac{x}{2N})}=\int^{\infty}_{\epsilon /N} \frac{dx}{x}\frac{1}{4\sinh^2(\frac{x}{2})}=\frac{N^2}{2\epsilon^2}+\frac{1}{12}\ln\frac{\epsilon}{N}+c_1+O(\epsilon)
\end{eqnarray}

Dependence on N of this integral completely comes from its dependence on $\epsilon$, which is easily established by differentiation on $\epsilon$ and expansion function under integral around $x=0$. Similar calculation for last contribution gives \cite{FL}:

\begin{eqnarray}
\int^{\infty}_\epsilon \frac{dx}{x}\left(-\frac{N^2}{e^x-1}\right)=-N^2\left(\frac{1}{\epsilon}+\frac{1}{2}\ln\epsilon +\frac{1}{2}(\gamma-\ln(2\pi))+O(\epsilon)\right)
\end{eqnarray}

The second contribution is:
\begin{eqnarray}
\int^{\infty}_\epsilon \frac{dx}{x}\frac{-e^{-x}}{4\sinh^2(\frac{x}{2N})}
\end{eqnarray}

Insert an expansion of $1/4\sinh^2(\frac{x}{2N})$:
\begin{eqnarray}
\int^{\infty}_\epsilon \frac{dx}{x}\frac{-e^{-x}}{4\sinh^2(\frac{x}{2N})}=\int^{\infty}_\epsilon \frac{dx}{x}e^{-x} \left( \frac{N}{2} \partial \coth(\frac{x}{2N})\right)=\\
\int^{\infty}_\epsilon \frac{dx}{x}e^{-x}\left(\frac{N}{2} \sum_{g=0}^{\infty} \frac{2B_{2g}(2g-1)}{(2g)!}\frac{x^{2g-2}}{N^{2g-1}}\right)=\\
-N^2\int^{\infty}_\epsilon dx\frac{e^{-x}}{x^3}+\int^{\infty}_\epsilon dx\frac{e^{-x}}{12x}+\sum_{g=2}^{\infty} \frac{B_{2g}}{2g(2g-2)}N^{2-2g} +O(\epsilon)
\end{eqnarray}
Evidently, for $g \geq 2$ asymptotic coincides with that of Barnes G-function, see (3.19) in \cite{Mar1}. For first two terms we have \cite{FL}:
\begin{eqnarray}
\int^{\infty}_\epsilon \frac{dx}{x^3}e^{-x}=\frac{1}{2\epsilon^2}-\frac{1}{\epsilon}-\frac{1}{2}\ln\epsilon+\frac{3}{4}-\frac{1}{2}\gamma+O(\epsilon)\\
\int^{\infty}_\epsilon\frac{dx}{x}e^{-x}=-\ln\epsilon-\gamma+O(\epsilon)
\end{eqnarray}

Summing up all contributions, we see that singular at $\epsilon \rightarrow 0$ terms cancel, other terms give asymptotic for $F_1$:

\begin{eqnarray} \label{nonpertF}
F_1(k)=-\ln Z_1=\frac{1}{2}(N^2-1)\ln \frac{k+N}{N}-F_2(0)=\\ \nonumber
\frac{1}{2}(N^2-1)\ln \frac{k+N}{N}+\frac{1}{12}\ln N +\frac{3}{4}N^2-\frac{1}{2}N^2\ln(2\pi)-c_1+\frac{1}{12}\gamma-\\ \nonumber
\sum_{g=2}^{\infty} \frac{B_{2g}}{2g(2g-2)}N^{2-2g} 
\end{eqnarray}

This coincides exactly with non-perturbative contribution given in (3.19) of \cite{Mar1} provided (taking into account $ \zeta'(-1) =1/12-\ln A $)

\begin{eqnarray}
c_1=\frac{1}{12}(\gamma+1)-\ln A
\end{eqnarray}
Here $ \gamma, A $ are Euler-Mascheroni and Glaisher constants, respectively. We check this equality up to sixth digit by numerical calculation of integral with "Mathematica", so we believe it is correct. One also have to take into account that (3.19) of \cite{Mar1}  includes a free energy of $U(1)$ theory, taken there to be $-(1/2)\ln (N/(k+N))$, so one have to add same quantity to our $SU(N)$ function (\ref{nonpertF}) to compare to (3.19).

So, for $g \geq 2$ we sum up two contributions into free energy, perturbative and non-perturbative:
\begin{eqnarray} \nonumber
\frac{2(2g-1)B_{2g}}{(2g)!(-1)} \int^{\infty}_0 dx \left( \frac{e^{-x}}{x(1-e^{-x})}\left(\frac{x}{\delta}\right)^{2g-2}\cosh\frac{Nx}{\delta}+\frac{1}{2x}\left(\frac{x}{\delta}\right)^{2g-2}e^{-\frac{xN}{\delta}}\right)=\\ 
\frac{2(2g-1)B_{2g}}{(2g)!(-1)} \int^{\infty}_0 dx  \frac{e^{-x/2}}{x(1-e^{-x})}\left(\frac{x}{\delta}\right)^{2g-2}(e^{x(-\frac{1}{2}+\frac{N}{\delta})}+e^{x(+\frac{1}{2}-\frac{N}{\delta})})=\\ \nonumber
\frac{2(2g-1)B_{2g}}{(2g)!(-1)}g_s^{2g-2} \int^{\infty}_0 dx  \frac{e^{-x/2}}{x(1-e^{-x})}\left(\frac{x}{2\pi}\right)^{2g-2}(e^{x(-\frac{1}{2}+\frac{\mu}{2\pi i})}+e^{x(+\frac{1}{2}-\frac{\mu}{2\pi i})})
\end{eqnarray}
and change integration variable $x\rightarrow 2\pi x$:

\begin{eqnarray} \label{Li}
\frac{2(2g-1)B_{2g}}{(2g)!(-1)}g_s^{2g-2} \int^{\infty}_0 dx  \frac{x^{2g-3}}{\sinh(\pi x)}\cosh(x(\pi-\frac{\mu}{i}))
\end{eqnarray}

Compare with the following integral representation of polylogarithm \cite{polylog}:

\begin{eqnarray}
Li_s(z)=\int^{\infty}_0 dx x^{-s} \frac{\sin\left(\frac{s\pi}{2}-x\ln(-z)\right)}{\sinh(\pi x)}
\end{eqnarray}
For $s=3-2g, g \geq 2$ integer
\begin{eqnarray}
Li_{3-2g}(z)=\int^{\infty}_0 dx x^{2g-3} \frac{\cos(x\ln(-z))}{\sinh(\pi x)}
\end{eqnarray}

\begin{eqnarray}
Li_{3-2g}(e^{-\mu})=\int^{\infty}_0 dx x^{2g-3} \frac{\cos(x(i\pi-\mu))}{\sinh(\pi x)}=\\
\int^{\infty}_0 dx x^{2g-3} \frac{\cosh(x(\pi-\frac{\mu}{i}))}{\sinh(\pi x)}
\end{eqnarray}
Comparing this with (\ref{Li}) we see that at $ g\geq 2 $ contribution into total free energy is 
\begin{eqnarray} \label{Li2}
\frac{2(2g-1)B_{2g}}{(2g)!(-1)}g_s^{2g-2}Li_{3-2g}(e^{-\mu})
\end{eqnarray}

in agreement with (5.32) of \cite{Mar1}.

\section{Barnes function and group volume function} \label{sectbarn}

As shown in previous Section, for $SU(N)$ group the asymptotics of our non-perturbative part of Chern-Simons theory coincides with previously known, essentially that of Barnes G-function. From the other side, these two functions are really different, since our one is even with respect to $N \rightarrow -N$ duality, but G function manifestly not. Now we shall suggest an explanation of this apparent contradiction. Consider integral representation of $F_1$:

\begin{eqnarray} 
F_1=\frac{1}{2}(N^2-1)\ln \frac{k+N}{N}-\int^{\infty}_0 \frac{dx}{x}\left(\frac{1-e^{-x}}{4\sinh^2(\frac{x}{2N})}-\frac{N^2}{e^x-1}\right)
\end{eqnarray}

This integral exist at all complex $N$ with non-zero real part . Let's try to connect points $N$ and $-N$ by some path in complex $N$ plane such that integral in $F_1$ make sense on all points of that path. It is easy to see that there is no such a path, because it inevitably has to pass through imaginary $N$ axis, but  integral is divergent at any point on that axis. From the other hand, any two points in semiplane with $\Re N>0$, or any two points with $\Re N<0$ can be connected by such a non-singular path. So we assume that our representation for non-perturbative free energy, and consequently for group volume defines  two analytic functions of $N$, one for positive $\Re N$, and one for negative $\Re N$. $1/N$ expansions of these analytical functions coincide with $G(1+N)$ and $G(1-N)$ in positive and negative $\Re N$ semiplanes, respectively.
The toy model for such a behavior is an integral

\begin{eqnarray}
\int_{0}^{\infty}\frac{dx}{\cosh(zx)}= \frac{\pi}{2z}, \Re z >0
\end{eqnarray}
which has features similar to our integral. It is not defined at $\Re z=0$, as our one for $\Re N=0$, it is symmetric w.r.t. the change $z\rightarrow -z $,  points $z, -z$ cannot be connected by continuous path with regular values of integral due to singular line $\Re z=0$. Calculation of integral gives an exact answer which actually provide an analytic continuation on all points on singular line except point z=0, which is true singularity. And in agreement with our assumption integral defines two analytic functions of z at $\Re z>0$ and $\Re z <0$: $\pi/2z$ and $-\pi/2z$, connected by $z\rightarrow -z $. 

We assume that at positive $\Re N$ our function, having the same $1/N$ asymptotic expansion and the same value at integer positive points (by construction) coincides essentially with $G(1+N)$. Exactly, this means that we get (new form of) the integral representation of logarithm of $G$-function (see \cite{Pe,OV,Mar1} for connection of $G$ and volume, for which we have an expression in above):

\begin{eqnarray} \label{GN}
\ln(G(1+N))=\frac{1}{2}N^2\ln(N)-\frac{1}{2}(N^2-N)\ln(2\pi)+\\ 
\int^{\infty}_0\frac{dx}{x}\left(\frac{1-e^{-x}}{4\sinh^2(\frac{x}{2N})}-\frac{N^2}{e^x-1}\right) \nonumber
\end{eqnarray}

Taking asymptotic expansion of integral from (\ref{nonpertF}):

\begin{eqnarray}
\int^{\infty}_0\frac{dx}{x}\left(\frac{1-e^{-x}}{4\sinh^2(\frac{x}{2N})}-\frac{N^2}{e^x-1}\right)= \\ \nonumber
-\frac{1}{12}\ln N -\frac{3}{4}N^2+\frac{1}{2}N^2\ln(2\pi)+c_1-\frac{1}{12}\gamma+\\ \nonumber
\sum_{g=2}^{\infty} \frac{B_{2g}}{2g(2g-2)}N^{2-2g} \label{inas}   
\end{eqnarray} 
we get an asymptotic expansion, which exactly coincides with that of $G(1+N)$, see e.g. \cite{Barnes1,GV2,OV}:  

\begin{eqnarray}
\ln(G(1+N))=\frac{1}{2}N^2\ln(N)-\frac{1}{2}(N^2-N)\ln(2\pi)+\\ \nonumber
-\frac{1}{12}\ln N -\frac{3}{4}N^2+\frac{1}{2}N^2\ln(2\pi)+c_1-\frac{1}{12}\gamma+\\ \nonumber
\sum_{g=2}^{\infty} \frac{B_{2g}}{2g(2g-2)}N^{2-2g} =\\   
(\frac{1}{2}N^2-\frac{1}{12})\ln(N)+\frac{1}{2}N\ln(2\pi)
 -\frac{3}{4}N^2+\frac{1}{12}-\ln A+\\
\sum_{g=2}^{\infty} \frac{B_{2g}}{2g(2g-2)}N^{2-2g}  \nonumber
\end{eqnarray}

This is the same coincidence which we observed earlier, comparing $1/N$ expansion of Chern-Simons and topological strings (\ref{nonpertF}). 

Next, for general point in Vogel's plane our volume function is natural group-theory based generalization of Barnes G-function (actually, it appears to be a combination of multiple (quadruple) Barnes' gamma functions with arguments linearly dependent from universal parameters, in preparation). The  analog of $N\rightarrow - N$ transformation is an exchange of parameters with different signs, and our integral representation, as in $SU(N)$ case, appears to be "glued" from few analytical functions, connected by permutation of parameters. One can calculate the corresponding generalization of well-known (Kinkelin) functional equation, connecting $G(1\pm N)$.

\section{Conclusion: possible directions of development} 

In this Section we shall  list possible directions of development of present results.

In their work \cite{GV2} Gopakumar and Vafa derive a structure result for free energy of topological string through so called Gopakumar-Vafa (integer) invariants. For the simplest case it corresponds to $SU(N)$ Chern-Simons theory on $S^3$ and is equal to 

\begin{eqnarray}
\sum_{m=1}^{\infty}\frac{ - e^{-m \mu}}{4m \sin^2(\frac{m g_s }{2})} \label{gp1}
\end{eqnarray}
plus $\mu$-independent terms given by (\ref{BB}). We shall recover below this result and show that actually some non-perturbative over string coupling constant terms are present.

Free energy is
 \begin{eqnarray}
 F_1+F_2=(dim/2)\ln(\delta/t) + \int^{\infty}_0 \frac{dx}{x} \frac{F(x/\delta)-F(x/t)}{(e^{x}-1)} 
 \end{eqnarray}
Let's extend integration to entire real $ x $ axis. Additional part is 

\begin{eqnarray}
 \int_{-\infty}^0 \frac{dx}{x} \frac{F(x/\delta)-F(x/t)}{(e^{x}-1)} = \int^{\infty}_0 \frac{dx}{x} \frac{F(x/\delta)-F(x/t)}{(1-e^{-x})}=\\
  \int^{\infty}_0 \frac{dx}{x} \frac{e^x (F(x/\delta)-F(x/t))}{(e^{x}-1)}=\\
   \int^{\infty}_0 \frac{dx}{x} \frac{e^x}{(e^{x}-1)} \left(\frac{\cosh(\frac{Nx}{k+N})-1}{2\sinh^2(\frac{x}{2(k+N)})}-\frac{\cosh(x)-1}{2\sinh^2(\frac{x}{2N})}\right)
 \end{eqnarray}
and is finite provided $ |N| < 1 $.

Difference is:
\begin{eqnarray}
 \int^{\infty}_0 \frac{dx}{x} (F(x/\delta)-F(x/t))
 \end{eqnarray}

This will be zero would integrals exist separately. But they diverge at upper limit $\Lambda $ as 

\begin{eqnarray}
 \int^{\Lambda}_0 \frac{dx}{x} F(x) = - N^2 \ln \Lambda +c +O(1/\Lambda)
 \end{eqnarray}
so difference is 
\begin{eqnarray}
  N^2 (\ln \delta/t) 
 \end{eqnarray}
so 
\begin{eqnarray}
 \int^{\infty}_{-\infty} = 2\int^{\infty}_0 + N^2 (\ln \delta/t)\\
 \int^{\infty}_0=\frac{1}{2}\int^{\infty}_{-\infty} -  (N^2/2) (\ln \delta/t)
 \end{eqnarray}
 
 So last term  almost cancels in total free energy, remaining part is an U(1) free energy:

\begin{eqnarray}
 F_1+F_2=\\ \nonumber
 \int_{-\infty}^{\infty} \frac{dx}{x} \frac{1}{(e^{x}-1)}\left(\frac{\sinh^2(\frac{Nx}{2(k+N)})}{\sinh^2(\frac{x}{2(k+N)})}-\frac{\sinh^2(\frac{x}{2})}{\sinh^2(\frac{x}{2N})}\right) -  (1/2) (\ln \delta/t)
\end{eqnarray}

As mentioned, this expression is applicable at $|N|<1$, so further results in principle are restricted to this (non-physical) domain. But it may be that further transformations will enlarge this domain of applicability, as happens in analytic continuation procedures, in all cases they deserve more careful further study. 

We would like to close contour of integration by upper semicircle, for further shrinking it to the poles in upper semiplane. It seems that it  is possible, due to analog of Jourdan's lemma. But we shall not discuss that issue now, since it evidently requires separate careful study, and simply assume that it is true, at least for some range of parameters, and look what will be the consequences. 

In that case we can shrink contour to poles in upper semiplane and integral becomes the sum of residues in these poles. Poles are at points 

\begin{eqnarray}
x=2\pi i p, \\ x=2\pi i(k+N) n \\ x=2\pi iN m
\end{eqnarray}
where $p,n,m $  are integers to be chosen so that corresponding pole be in upper semiplane. For real positive $k$ and $N$ all positive $p,n,m$ should be taken. Remind that there is no pole at $x=0$. Contribution of three series of poles are (assuming there is no coincidence between them):

\begin{eqnarray} \label{gp1-1}
\sum_{m=1}^{\infty}\frac{ - \cos(\frac{m N 2\pi }{\delta})+1}{2m \sin^2(\frac{m \pi }{\delta}) }=
\sum_{m=1}^{\infty}\frac{1 - \cosh(m \mu)}{2m \sin^2(\frac{m g_s }{2})}
\end{eqnarray}

\begin{eqnarray}
\sum_{m=1}^{\infty}i\frac{-1+e^{-2\pi iNm}(1+2\pi i N m)}{2m^2\pi}
\end{eqnarray}

\begin{eqnarray}
\sum_{m=1}^{\infty} \left( \frac{m N \pi \sin(2m N \pi)- \sin^2(m N \pi ) }{m^2 \pi ^2 \left(e^{2m i \delta \pi} -1\right)}-
2 i \delta \frac{\sin^2(m N \pi ) e^{2m i \pi \delta  }  }{m \pi  \left(e^{2m i \delta \pi} -1\right)^2} \right)
\end{eqnarray}

Actually (\ref{gp1-1}) coincides with (\ref{gp1}) plus (\ref{BB}), taking into account the invariance of given polylogarithm  $Li_{3-2g}(e^{-\mu})=Li_{3-2g}(e^{\mu})$ in (\ref{Li2}).  Next contributions are  non-perturbative w.r.t. the string coupling constant $g_s=2\pi/\delta$.  They have to be compared with existing non-perturbative calculations, see particularly \cite{PS}.  

Evidently, these calculations have to be clarified and justified in many respects, we hope to do that elsewhere. 

Universal formulation of Chern-Simons theory open the way for the generalizations of many faces of that theory. One possibility is an extension of $1/N$ expansion on exceptional groups. In some sense it is already achieved. Indeed, in Vogel's plane all five exceptional groups are located on the line (usually denoted Exc) $\gamma=2(\alpha+\beta)$, just as e.g. $SU(N)$ are located on line $\alpha+\beta=0$. One can introduce any linear parameter on the line Exc, say $z=\alpha/\beta$ (for $SU(N)$ standard parameter is $N=2\gamma/\beta$) and expand all universal quantities (partition function, adjoint Wilson loop, etc.) in a (Laurent) series w.r.t. the parameter $z$ or $z^{-1}$. But of course it is not enough, we would like to have a topological interpretation of corresponding coefficients, just as for $SU(N)$ (and $SO/Sp$) Ooguri and Vafa \cite{OV} interpret coefficients in $1/N$ expansion as string theory objects, namely as a (virtual) Euler characteristics of moduli space of surfaces of a given genus, which are proportional to Bernoulli numbers, (\ref{nonpertF}). In Exc line case, as in Section \ref{secsun} for $SU(N)$, to get an expansion of free energy over parameter $z$ one should substitute into expression for universal character (\ref{gene}) universal parameters as a (linear) functions of $z$ and expand that over $z$. Difference appear first in that this object is not polynomial, or more exactly it becomes polynomial when common multiplier  $dim \, \mathfrak {g}$ (dimension of algebra) is separated. Second, corresponding expansion includes not Bernoulli numbers, but similar objects relevant for Barnes' multiple gamma-functions  \cite{Barnes2,Barnes3,Rui}.

Other possible generalization is substitution of  $S^3$ manifolds by some other three-dimensional manifold. It seems to be possible to obtain  universal expressions  for partition function on other manifolds provided one obtain a universal representation for some other (than $S_{00}$) elements of S-matrix of modular transformations, since partition functions for other manifolds can be obtained by topological manipulations with the use of that matrix elements \cite{W1}.  

Finally, this view from universality side suggests some duality between topological string's coupling constant $g_s$ and K\"ahler  parameter(s) $\mu$. Indeed, initial universal functions (for free energy, e.g.) are symmetric w.r.t. the universal parameters. When restricted to $SU(N)$ some of these parameters become string coupling constant, namely $\alpha/\delta \sim g_s, \beta/\delta\sim g_s$, third one becomes K\"ahler parameter: $\gamma/\delta \sim \mu$, which naturally hints on some hidden duality between these parameters of topological string.

Note also recent work \cite{KS} where integral representation \ref{totalfree} of free energy is extended on refined Chern-Simons theory on $S^3$.

\section{Acknowledgments.}

I'm indebted to A.Zamolodchikov for discussions and for advising a Borel transform, and to H.Khudaverdian, H.Mkrtchyan and A.Veselov for interest to this work.  I'm also indebted to Simons Center for Geometry and Physics, where this work started, and to organizers of Integrability Program N.Nekrasov and S.Shatashvili for invitation. Work is partially supported by the grant of the Science Committee of the Ministry of Science and Education of the Republic of Armenia under contract 11-1c037, and grant  of Volkswagen Foundation.


\begin{thebibliography}{99}

\bibitem{H1}
G. 't Hooft, {\it A planar diagram theory for strong interactions.} Nucl.Phys. {\bf B72} (1974), 461-473.

\bibitem{Mkr}
R.L. Mkrtchyan, {\it The equivalence of $Sp(2N)$ and $SO(-2N)$ gauge theories.} Phys. Lett. {\bf 105B} (1981), 174-176.

\bibitem{Cvitbook}
P. Cvitanovic, {\it Group Theory.} Princeton Univ. Press, Princeton, NJ, 2004. http://www.nbi.dk/group theory

\bibitem{MV1}
R.L. Mkrtchyan, A.P. Veselov, {\it Universality in Chern-Simons theory}, JHEP 08 (2012) 153, arXiv:1203.0766.

\bibitem{V0}
P.Vogel, {\it Algebraic structures on modules of diagrams.}, Preprint (1995),  
J. Pure Appl. Algebra {\bf 215} (2011), no. 6, 1292-1339.

\bibitem{V}
 P. Vogel, {\it The universal Lie algebra}, Preprint (1999).
 
\bibitem{Del}
P. Deligne, {\it La s\'erie exceptionnelle des groupes de Lie.} C. R. Acad. Sci. Paris, S\'erie I {\bf 322} (1996), 321-326.

\bibitem {DM}
P. Deligne and R. de Man, {\it La s\'erie exceptionnelle des groupes de Lie II.} C. R. Acad. Sci. Paris, S\'erie I  {\bf 323} (1996), 577-582.  
 
\bibitem{LM1} 
J.M. Landsberg, L. Manivel, {\it A universal dimension formula for complex simple Lie algebras.} Adv. Math. {\bf 201} (2006), 379-407

\bibitem{LM2}
J. M. Landsberg, L. Manivel, {\it Triality, exceptional Lie algebras and Deligne dimension formulas.} Adv. Math. {\bf 171} (2002), 59-85.

\bibitem{LM4} 
J. M. Landsberg, L. Manivel, {\it{Series of Lie groups}}, Michigan Math. J. (2004), 52(2): 453-479,
MR2069810.

\bibitem{MSV}
R.L. Mkrtchyan, A.N. Sergeev, A.P. Veselov {\it Casimir eigenvalues for universal Lie algebra.} Journ. Math.Phys. 53, 102106 (2012), arXiv:1105.0115 (2011).

\bibitem{Mkr2}
R.L.Mkrtchyan,{\it On a map of Vogel's plane}, arxiv:1209.5709.

\bibitem{Pe} 
 V. Periwal {\it Topological closed string interpretation of Chern-Simons theory.} Phys. Rev. Lett. {\bf 71} (1993), 1295; hep-th/9305115.

\bibitem{GV} R. Gopakumar, C. Vafa, {\it On the Gauge Theory/Geometry Correspondence.} Adv.Theor.Math.Phys. {\bf 3} (1999) 1415-1443, arXiv:hep-th/9811131v1. 

\bibitem{OV}
H. Ooguri and C. Vafa {\it Worldsheet derivation of a large N duality.} Nucl. Phys. {\bf B 641} (2003), hep-th/0205297.

\bibitem{Mar1} 
Marcos Mari\~no, {\it Chern-Simons Theory and Topological Strings}, Rev.Mod.Phys.77:675-720,2005, 
arXiv:hep-th/0406005

\bibitem{W1} 
 E. Witten, {\it Quantum field theory and the Jones polynomial.} Comm. Math. Phys. {\bf 121} (1989),  351-399.
 
\bibitem{gamma} 
http://functions.wolfram.com/06.11.07.0004.01

\bibitem{KP}
V. G. Kac,  D. H. Peterson, {\it Infinite-dimensional Lie algebras, theta functions and modular forms}, Advances in  Mathematics 53 (1984), 125-264.

\bibitem{Vv}
 A.P.Veselov, communication at April 2013

\bibitem{GV1}
Rajesh Gopakumar, Cumrun Vafa, {\it M-Theory and Topological Strings-I},  arXiv:hep-th/9809187

\bibitem{GV2}
Rajesh Gopakumar, Cumrun Vafa, {\it M-Theory and Topological Strings-II},  arXiv:hep-th/9812127

\bibitem{Barnes1}
E.W.Barnes, {\it The theory of the G-function}, Quarterly Journ. Pure and Appl. Math. 31 (1900), 264–314.

\bibitem{FL} 
Chelo Ferreira, Jose L. Lopez, {\it An Asymptotic Expansion of the Double Gamma Function}, Journal of Approximation Theory, Volume 111, Issue 2, 2001, pp. 298-314

\bibitem{polylog}
http://en.wikipedia.org/wiki/Polylogarithm

\bibitem{PS}
Sara Pasquetti and Ricardo Schiappa, {\it and Stokes Nonperturbative Phenomena in Topological String Theory and c=1 Matrix Models}, arXiv:0907.4082.

\bibitem{Barnes2}
E. W. Barnes,{\it The Theory of the Double Gamma Function}, Philosophical Transactions of the Royal Society of London. Series A, 
Vol. 196, (1901), pp. 265-387

\bibitem{Barnes3}
E.W. Barnes, {\it On the theory of the multiple gamma function}. Trans. Cambridge Philos. Soc.
19 (1904), 374-425

\bibitem{Rui}
S. N. M. Ruijsenaars, {\it On Barnes' Multiple Zeta and Gamma Functions}, Advances in Mathematics 156, 107-132 (2000)

\bibitem{KS}
Daniel Krefl and Albert Schwarz,  {\it Refined Chern-Simons versus Vogel universality}, arXiv:1304.7873 [hep-th]

\end{thebibliography}
\end{document}